%% file: 00main.tex
\documentclass[twoside,twocolumn,9pt]{article}
\usepackage{extsizes}
\usepackage[super,sort&compress,comma]{natbib} 
\usepackage[version=3]{mhchem}
\usepackage[left=1.5cm, right=1.5cm, top=1.785cm, bottom=2.0cm]{geometry}
\usepackage{balance}
\usepackage{mathptmx}
\usepackage{sectsty}
\usepackage{graphicx} 
\usepackage{lastpage}
\usepackage[format=plain,justification=justified,singlelinecheck=false,font={stretch=1.125,small,sf},labelfont=bf,labelsep=space]{caption}
\usepackage{float}
\usepackage{fancyhdr}
\usepackage{fnpos}
\usepackage[english]{babel}
\usepackage{array}
\usepackage{droidsans}
\usepackage{charter}
\usepackage[T1]{fontenc}
\usepackage[usenames,dvipsnames]{xcolor}
\usepackage{setspace}
\usepackage[compact]{titlesec}
\usepackage{hyperref}
\usepackage{gensymb}
\usepackage{upgreek}
\usepackage{epstopdf}%This line makes .pdf figures into .pdf - please comment out if not required.
\usepackage{xcolor}
\usepackage{float}

\definecolor{cream}{RGB}{222,217,201}

\begin{document}

%%% Style
% {
\pagestyle{fancy}
\thispagestyle{plain}
\fancypagestyle{plain}{
%%%HEADER%%%
\renewcommand{\headrulewidth}{0pt}
}
%%%END OF HEADER%%%

%%%PAGE SETUP - Please do not change any commands within this section%%%
\makeFNbottom
\makeatletter
\renewcommand\LARGE{\@setfontsize\LARGE{15pt}{17}}
\renewcommand\Large{\@setfontsize\Large{12pt}{14}}
\renewcommand\large{\@setfontsize\large{10pt}{12}}
\renewcommand\footnotesize{\@setfontsize\footnotesize{7pt}{10}}
\makeatother

\renewcommand{\thefootnote}{\fnsymbol{footnote}}
\renewcommand\footnoterule{\vspace*{1pt}% 
\color{cream}\hrule width 3.5in height 0.4pt \color{black}\vspace*{5pt}} 
\setcounter{secnumdepth}{5}

\makeatletter 
\renewcommand\@biblabel[1]{#1}            
\renewcommand\@makefntext[1]% 
{\noindent\makebox[0pt][r]{\@thefnmark\,}#1}
\makeatother 
\renewcommand{\figurename}{\small{Fig.}~}
\sectionfont{\sffamily\Large}
\subsectionfont{\normalsize}
\subsubsectionfont{\bf}
\setstretch{1.125} %In particular, please do not alter this line.
\setlength{\skip\footins}{0.8cm}
\setlength{\footnotesep}{0.25cm}
\setlength{\jot}{10pt}
\titlespacing*{\section}{0pt}{4pt}{4pt}
\titlespacing*{\subsection}{0pt}{15pt}{1pt}
%%%END OF PAGE SETUP%%%

%%%FOOTER%%%
\fancyfoot{}
%\fancyfoot[LO,RE]{\vspace{-7.1pt}\includegraphics[height=9pt]{head_foot/LF}}
%\fancyfoot[CO]{\vspace{-7.1pt}\hspace{13.2cm}\includegraphics{head_foot/RF}}
%\fancyfoot[CE]{\vspace{-7.2pt}\hspace{-14.2cm}\includegraphics{head_foot/RF}}
\fancyfoot[RO]{\footnotesize{\sffamily{ ~\textbar  \hspace{2pt}\thepage}}}
\fancyfoot[LE]{\footnotesize{\sffamily{\thepage~\textbar}}}
\fancyhead{}
\renewcommand{\headrulewidth}{0pt} 
\renewcommand{\footrulewidth}{0pt}
\setlength{\arrayrulewidth}{1pt}
\setlength{\columnsep}{6.5mm}
\setlength\bibsep{1pt}
%%%END OF FOOTER%%%

%%%FIGURE SETUP - please do not change any commands within this section%%%
\makeatletter 
\newlength{\figrulesep} 
\setlength{\figrulesep}{0.5\textfloatsep} 

\newcommand{\topfigrule}{\vspace*{-1pt}% 
\noindent{\color{cream}\rule[-\figrulesep]{\columnwidth}{1.5pt}} }

\newcommand{\botfigrule}{\vspace*{-2pt}% 
\noindent{\color{cream}\rule[\figrulesep]{\columnwidth}{1.5pt}} }

\newcommand{\dblfigrule}{\vspace*{-1pt}% 
\noindent{\color{cream}\rule[-\figrulesep]{\textwidth}{1.5pt}} }

\makeatother
%%%END OF FIGURE SETUP%%%
% }

%%%TITLE, AUTHORS AND ABSTRACT%%%
% {
\twocolumn[
  \begin{@twocolumnfalse}
%{\includegraphics[height=30pt]{head_foot/SM}\hfill\raisebox{0pt}[0pt][0pt]{\includegraphics[height=55pt]{head_foot/RSC_LOGO_CMYK}}\\[1ex]
%\includegraphics[width=18.5cm]{head_foot/header_bar}}\par
\vspace{8em}
\sffamily
\begin{tabular}{m{4.5cm} p{13.5cm} }

 & \noindent\LARGE{\textbf{%
3D Visualization Reveals the Cooling Rate Dependent Crystallization near a Wall in Dense Microgel Systems$^\dag$
}}
 \\

\vspace{0.3cm} & \vspace{0.3cm} \\

 & \noindent\large{M.P.M. Schelling, T.W.J. Verouden, T.C.M. Stevens, and J.-M. Meijer$^{\ast}$} \\%Author names go here instead of "Full name", etc.

& \noindent\normalsize{

Controlled crystallization, melting and vitrification are important fundamental processes in nature and technology. However, the microscopic details of these fundamental phenomena still lack understanding, in particular how the cooling rate and presence of a wall influence the crystal nucleation and glass formation. Thermoresponsive microgels provide the possibility to study phase transitions on a single-particle level, owing to the ability to tune the particle size with temperature. In this study, we employ composite microgels consisting of a hard core and a crosslinked poly(\textit{N}-isopropyl acrylamide-\textit{co}-methacrylic acid) shell to study the crystallization of dense suspensions of soft colloids near a wall using confocal microscopy. We investigate the effect of cooling rate on the fluid-to-solid transition close to the sample wall. The structures formed during cooling range from glassy in case of a rapid temperature quench, to crystalline when a slow cooling rate is used. Detailed analysis of the final structure reveals that the cooling rate also sets the degree of alignment of the crystal domains with the wall as a result of a balance between homogeneous and heterogenous crystal nucleation. The results presented here yield valuable insight into the microscopic details of temperature-controlled crystallization near a wall. This understanding will help pave the way towards optimal crystallization conditions for microgel applications.
}

\end{tabular}

 \end{@twocolumnfalse} \vspace{0.6cm}
]
%%%END OF TITLE, AUTHORS AND ABSTRACT%%%

%%%FONT SETUP - please do not change any commands within this section
\renewcommand*\rmdefault{bch}\normalfont\upshape
\rmfamily
\section*{}
\vspace{-1cm}

%%%FOOTNOTES%%%

\footnotetext{Department of Applied Physics and Science Education, Eindhoven University of Technology, Groene Loper 19, 5612 AP Eindhoven, The Netherlands.}
\footnotetext{Institute for Complex Molecular Systems, Eindhoven University of Technology, Groene Loper 19, 5612 AP Eindhoven, The Netherlands.}
\footnotetext{$^{\ast}$~E-mail: j.m.meijer@tue.nl}

%Please use \dag to cite the ESI in the main text of the article.
%If you article does not have ESI please remove the the \dag symbol from the title and the footnotetext below.
\footnotetext{\dag~Electronic Supplementary Information (ESI) available: Details on the microgel synthesis and characterization, determination of the melting point of the suspension, and additional figures about the temperature-controlled experiments and subsequent analysis are presented.}
%additional addresses can be cited as above using the lower-case letters, c, d, e... If all authors are from the same address, no letter is required

%%%END OF FOOTNOTES%%%
% }

%%%MAIN TEXT%%%%

\section{Introduction}

Microgels are soft and deformable colloidal particles consisting of a polymer network swollen by a liquid. On the one hand, microgels behave similar to hard-like monodisperse colloids, as they are able to crystallize when dispersed at a sufficiently high particle density \cite{Senff1999TemperatureSpheres, Lyon2004MicrogelCrystals, Mohanty2008StructuralMicrogels}. On the other hand, microgels exhibit macromolecule-like properties that are of particular importance in dense systems, for instance deformation, compression, and interpenetration/entanglement of dangling chains \cite{Plamper2017FunctionalSystems, Mohanty2017InterpenetrationDensities, Conley2017JammingCompression, Conley2019RelationshipMicrogels, Nikolov2020BehaviorSuspensions}. Additionally, microgels exhibit a responsiveness to external conditions such as temperature, pH and ionic strength that can be tuned via their chemical composition and crosslink-density \cite{Karg2019NanogelsTrends}. For this reason, responsive microgels have played a key role as intriguing model system for studying the phase behavior of dense systems of soft colloids \cite{Yunker2014PhysicsParticles}.

The most widely investigated microgels are those prepared from poly(\textit{N}-isopropyl acrylamide) (PNIPAM) \cite{Karg2019NanogelsTrends, Pelton1986PreparationN-isopropylacrylamide, Pelton2000Temperature-sensitiveMicrogels}. PNIPAM exhibits a Lower Critical Solution Temperature (LCST) in water above which its solubility sharply decreases, resulting in a temperature-controlled size change of microgels prepared from PNIPAM. At low temperatures, PNIPAM microgels are swollen with water, while the microgels collapse when the suspension is heated above the LCST. For PNIPAM microgels this Volume Phase Transition can be up to 90\% in volume\cite{Pelton2000Temperature-sensitiveMicrogels} and is typically around 32 \degree C (the Volume Phase Transition Temperature, or VPTT). Due to this size-tunability, the volume fraction of PNIPAM microgel suspensions can be controlled via temperature, which allows to reversibly transition between a fluid phase at low volume fractions (high temperature) and a solid phase at high volume fractions (low temperature). Composite microgels with a hard core-PNIPAM shell morphology are of particular interest for studying dense systems, as the presence of small cores can increase the contrast in scattering experiments \cite{Hildebrandt2022SAXSFraction, Hildebrandt2023FluidsolidMicrogels}, or help distinguishing microgels in confocal microscopy experiments \cite{Higler2013SubstitutionalCrystals, Appel2015TemperatureSuspensions}. As a result PNIPAM microgels have been a popular tool in experimental studies on crystallization \cite{Meng2007CrystallizationMicrogels, Bocanegra-Flores2022CrystallizationFractions, Ganapathi2021StructureGlass}, melting \cite{Wang2016MeltingCrystals, Alsayed2005PremeltingCrystals}, and the glass transition \cite{Frenzel2021Glass-liquidParticles}. 

\begin{figure}[t!]
\centering
  \includegraphics[]{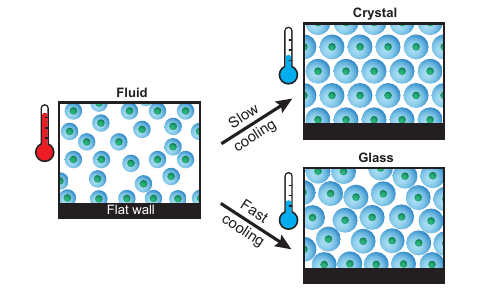}
  \caption{Schematic representation of the non-responsive core - PNIPAM shell microgels used in this work. Depending on the cooling rate, the microgels can form a dense crystalline or glassy structure. Due to their surface active nature, a layer of adsorbed microgels is present on the substrate wall.}
  \label{Figure 1}
\end{figure}

For dense PNIPAM microgel suspensions it is well known that, similar to molecular liquids and metals, the temperature change influences the final state that forms, leading to either crystals or glassy structures (Fig. \ref{Figure 1}). For metals, control over the exact cooling rate has been shown to tune the degree of homogeneous versus heterogeneous crystal nucleation \cite{Mittemeijer2021FundamentalsScience, Kalikmanov2013NucleationTheory}. However, for PNIPAM microgels so far the influence of the exact cooling rate on the crystallization process has received only limited attention, in contrast to fixed temperature experiments \cite{Meng2007CrystallizationMicrogels, Tang2004CrystallizationSpectroscopy, Bocanegra-Flores2022CrystallizationFractions, Lyon2004MicrogelCrystals}. In a recent study Lapkin et al.\cite{Lapkin2022InScattering} employed small-angle x-ray scattering to study the crystallization and melting of dense PNIPAM suspensions. The authors showed that for a slow cooling rate a large single crystal domain formed. During the formation of the crystal domain different types of stacking disorder were found to develop at different rates. In addition, upon heating the crystal domain `inhomogenous' melting was observed indicating the presence of structural heterogeneity within the crystal domain. However, which exact structural differences were present and how the cooling rate influenced these was not clear. 

The most likely cause of these structural differences in the dense PNIPAM microgel system, is the presence of the sample chamber wall. A substrate wall is known to have an important effect on the crystal structures formed by colloidal particles \cite{Sandomirski2014HeterogeneousWalls, Arai2017Surface-assistedColloids}. For instance, the presence of a wall has been shown in hard sphere colloids to lower the barrier for crystal nucleation \cite{Auer2003LineColloids} and cause the alignment of hexagonally-packed planes \cite{Auer2003LineColloids, Hoogenboom2003AWall, Sandomirski2011HeterogeneousWall}. In addition, the structural features of a wall, such as curvature \cite{Cacciuto2004OnsetSuspensions}, topography \cite{Mishra2016Site-specificTransport} and shear effects \cite{Orr2022Single-orientationShear}, also play an important role. By patterning the wall, it was shown that the wall can even act as a template to control the structure and orientation of the crystal (colloidal epitaxy) \cite{vanBlaaderen2003ColloidalCrystallization, Cacciuto2005SimulationTemplates}. In experiments with PNIPAM microgels, due to their surface active nature \cite{Plamper2017FunctionalSystems}, the glass wall of a sample cell is usually covered by a disordered layer of strongly adsorbed microgels. Clearly, wall effects can often not be neglected when studying phase transitions with experiments. So far, however, the influence of the sample wall, and in particular combined with cooling rate effects, on the crystallization of microgels remains elusive. 

In this work, we investigate the crystallization near a wall in a dense microgel suspension using various rates of continuous cooling with temperature-controlled Confocal Laser Scanning Microscopy (CLSM) experiments. To image and locate the individual microgels even in a very dense system, we employed composite microgels that contain a fluorescent, non-responsive core and a non-fluorescent, thermoresponsive PNIPAM shell \cite{Appel2015TemperatureSuspensions}. Fast and precise control over the temperature was achieved by using a temperature-controllable VAHEAT substrate \cite{Icha2022PreciseVAHEAT} and allows us to apply different cooling rates. We find that the system forms solid phases ranging from those with a high degree of crystallinity to glass-like phases. We further find that the wall influences the crystal orientation depending on the cooling rate applied. The results presented here provide detailed information on a single-particle level about the effect of the cooling rate on the ordering in PNIPAM microgel suspensions, and underline the importance of wall-effects on the final crystal structure.

\section{Experimental methods}
\subsection{Synthesis and characterization of PNIPAM microgels}
Thermoresponsive microgels consisting of poly(2,2,2-trifluoroethyl methacrylate) cores and PNIPAM shells were synthesized according to the procedure by Appel et al. \cite{Appel2015TemperatureSuspensions}. The non-responsive cores have a hydrodynamic diameter $D_{\mathrm{h}} = 0.192 \pm 0.002 ~\upmu \textrm{m}$ and contain Pyrromethene 546 (BODIPY) dye to allow for fluorescent imaging. The PNIPAM shell contains 1.0 mol\% crosslinker (\textit{N},\textit{N}'-methylenebisacrylamide) and 6.9 mol\% methacrylic acid.  Details about the synthesis can be found in Section S1 of the ESI\dag. The microgels have a hydrodynamic diameter of $D_{\mathrm{h}}(20 ~\degree \mathrm{C}) = 1.04 \pm 0.02 ~\upmu \textrm{m}$ in swollen state (measured in 10 mM \ce{NaCl}); see also the Supplementary Fig. 1\dag ~for a complete overview of the DLS results.
A microgel suspension with a concentration of 7 wt\% in water at 10 mM \ce{NaCl} was used for all experiments, which corresponds to a number density of $\rho_{0} = 2.5 ~\upmu \textrm{m}^{-3}$ as determined with CLSM by counting microgels within a given volume. Due to ability of microgels to deform and deswell at high densities, it is problematic to determine the \textit{true} volume fraction of a suspension. Instead, we estimate the \textit{effective} volume fraction from $D_{\mathrm{h}}$ (measured in dilute conditions) via $\phi_{\mathrm{eff}} = \rho_{0} \frac{\pi}{6} D_{\mathrm{h}}^{3}$ and find for the temperature range used in our experiments a decrease from $\phi_{\mathrm{eff}}(20 ~\degree \mathrm{C}) = 1.46$ to $\phi_{\mathrm{eff}}(28 ~\degree \mathrm{C}) = 1.27$. Note that $\phi_{\mathrm{eff}} > 1$ indicates a substantial amount of deswelling or interpenetration of the microgels.

\subsection{CLSM experiments}
Temperature-controlled CLSM experiments were performed using a Nikon A1R HD25 microscope equipped with a 100x oil immersion objective (Nikon CFI Plan Apo VC, NA = 1.4), 488 nm laser and GaAsP PMT detector. The sample temperature was controlled using a VAHEAT (Interherence) controller. Sample cells were prepared by gluing a glass ring (5 mm inner diameter) to a VAHEAT substrate. On top of the ring, a coverslip with two holes was glued, through which the cell was filled with the microgel suspension. The sample cell was sealed air-tight with a second coverslip that was glued on top to cover the holes. More details about the sample cell can be found in Section S2 of the ESI\dag.

The sample was cooled from 28.0 to 20.0 \degree C using rates of 0.1 \degree C/min and 0.5 \degree C/min, and by reducing the temperature in a single step (i.e. a rapid temperature quench), all repeated three times. Before each cooling ramp, the sample was kept at 28 \degree C for 60 s for temperature equilibration. CLSM xyzt-scans were obtained using a voxel size of 0.063x0.063x0.100 $\upmu \textrm{m}$ and an image size of 1024x1024x161 (acquisition time is approximately 80 s) above the coverslip for a total of 2 h after start of the cooling ramps. Axial distances were corrected for the refractive index mismatch between water and the immersion oil \cite{Besseling2015MethodsIndex}. The final volume near the coverslip that is investigated is approximately 64x64x10 $\upmu \textrm{m}$, containing around $10^5$ microgels. All xyzt-scans were deconvolved in NIS-Elements AR software before analysis.

\subsection{Particle tracking and structural analysis}
Particle tracking was performed using Trackpy \cite{Allan2021Soft-matter/trackpy:v0.5.0} based on the Crocker-Grier centroid finding algorithm \cite{Crocker1996MethodsStudies}. Only microgels in a crystal or glassy state are located accurately as microgels in the fluid phase (i.e. at higher temperatures) typically move too fast for 3D imaging. Structural analysis was performed using Freud \cite{Ramasubramani2020Freud:Data}, OVITO \cite{Stukowski2010VisualizationTool}, and in-house Python scripts.

\begin{figure}[b!]
\centering
  \includegraphics[]{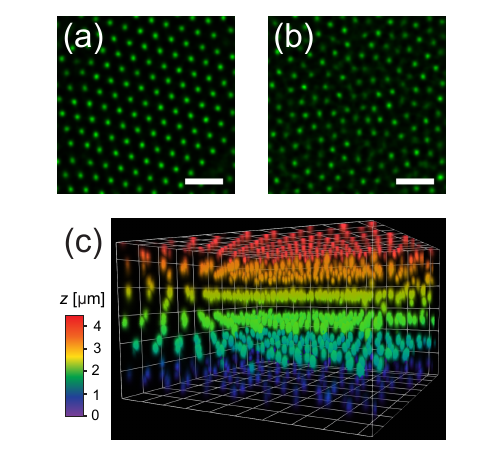}
  \caption{Deconvolved CLSM images at 20 \degree C of the dense microgel suspension; in (b) a crystalline structure and in (c) a glassy structure (slices of 3D stack). Note that only the cores are visible. Scale bars are 2 \textmu m. (c) Reconstructed intensity of 3D CLSM stack with dimensions 9.4x9.4x4.5$~\upmu \textrm{m}$.}
  \label{Figure 2}
\end{figure}

\section{Results and discussion}

\subsection{Phase behavior and structure at the wall}
The dense microgel system investigated in this work consists of a suspension of weakly-crosslinked PNIPAM microgels with a fluorescent core \cite{Appel2015TemperatureSuspensions} that is in contact with a horizontal, temperature-controlled wall, i.e. a VAHEAT substrate. The microgels are densily packed ($\phi_{\mathrm{eff}}(20 ~\degree \mathrm{C}) = 1.46$), and the suspension is in an arrested state at 20 \degree C. Heating the suspension to 28 \degree C results in a transition to a fluid state (see ESI\dag, Section S3), even though the effective volume fraction $\phi_{\mathrm{eff}}$ remains high ($\phi_{\mathrm{eff}}(28 ~\degree \mathrm{C}) = 1.27$). It should be noted here, however, that deswelling in dense systems \cite{Scotti2016TheSuspensions} results a \textit{true} volume fraction much lower than $\phi_{\mathrm{eff}}$, but is experimentally difficult to determine. At 20 \degree C, dependent on the cooling rate applied to the sample from the fluid phase, the microgels assemble in either a crystalline or glassy structure. Typical CLSM images of the crystal and glass phases are shown in Figs. \ref{Figure 2}(a)-(b), respectively. In addition, a 3D rendering of the fluorescent cores in a part of the xyzt-scan is shown in Fig. \ref{Figure 2}(c). These CLSM images confirm that each microgel core position can be visualized with high accuracy allowing detection of the microgels during the phase transitions, even when the effective volume fraction is in a regime where interpenetration and deformation of the microgels typically plays a role. 

\begin{figure}[b]
\centering
  \includegraphics[width = 1\columnwidth]{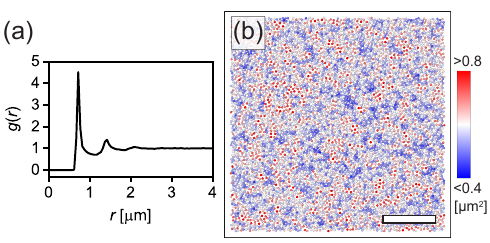}
  \caption{(a) 2D radial distribution function $g(r)$ of microgels adsorbed to the coverslip.  (b) Typical reconstruction of adsorbed microgels; the color indicates the Voronoi cell area. Scale bar is 15 $\upmu \textrm{m}$.}
  \label{Figure 3}
\end{figure}

It is well known that PNIPAM microgels display surfactant-like behavior \cite{Plamper2017FunctionalSystems} which results in strong adsorption onto interfaces \cite{Wellert2015ResponsiveInterfaces}. Indeed, we find that the horizontal glass coverslip in the system considered here is covered with immobile microgels with no long-range order. We also note that the microgels remain irreversibly stuck to the coverslip during the all experiments and the temperature range (20 -- 28 \degree C) used in this work. To confirm the absence of order, we determined the particle positions of the absorbed layer of microgels on the wall. Figure \ref{Figure 3}(a) shows a typical radial distribution function $g(r)$ for the microgels at the wall. The $g(r)$ shows only few peaks confirming the absence of any long-range order in the densely packed layer. We find a mean distance between nearest neighbors of $0.73 ~\upmu \textrm{m}$, much smaller than the hydrodynamic diameter of the microgels, due to deswelling/deformation of the microgels. Next, we investigated whether any spatial variation exists in packing of adsorbed microgels using Voronoi tessellation. In short, this method assigns to each adsorbed microgel a Voronoi cell that consists of all points closest to that microgel. Figure \ref{Figure 3}(b) shows the areas of the Voronoi cells for all detected microgels in a typical field-of-view. Here, short-ranged spatial flucuations can be observed that indicate small local differences in the packing density of adsorbed microgels. Clearly, the packing of the microgels on the wall appears to be random and is thus expected not to promote crystallization.

\subsection{Effect of cooling rate} \label{Effect of cooling rate}
\begin{figure}[h!]
 \centering
 \includegraphics[]{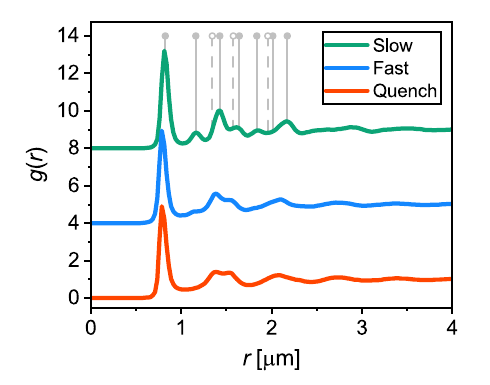}
 \caption{Average 3D radial distribution functions $g(r)$ of microgels in area 10 \textmu m above the coverslip at the end (2 h) of the three different cooling ramps. Curves are offset for clarity. Positions of the first peaks are $0.82 \pm 0.01 ~\upmu \textrm{m}$ (slow), $0.80 \pm 0.01 ~\upmu \textrm{m}$ (fast), and $0.80 \pm 0.01 ~\upmu \textrm{m}$ (quench).  Vertical grey lines indicate the expected peak positions for an FCC (solid lines) and HCP crystal (solid and dashed lines).}
 \label{Figure 4}
\end{figure}

To investigate the effect of cooling rate on the crystallization of the microgel system, we performed temperature-controlled CLSM experiments by cooling down from 28.0 \degree C to 20.0 \degree C with different cooling ramps, all repeated three times. The cooling rates used were: 0.1 \degree C/min (slow), 0.5 \degree C/min (fast), and a step-like temperature drop (quench); we will refer to the different cooling rates in the rest of the manuscript as slow, fast and quench as indicated between the parentheses. For the quench, the sample temperature dropped below 21.0 \degree C within the first 30 seconds after the the start of the ramp, which corresponds to a cooling rate of >10 \degree C/min within that time (see Supplementary Fig. 4 in the ESI\dag ~for a comparison between the measured and setpoint temperature). During the cooling ramps, CLSM xyzt-scans were obtained to capture the particles positions in a region approximately 10 $\upmu \textrm{m}$ above the coverslip. From the xyzt-stacks we extracted all particle positions in solid structures, i.e. crystalline or glassy, as the particles in a fluid diffuse too fast for accurate capture during a single xyz-scan. First, we analyzed the overall order in the final dense phases obtained after the cooling ramps to identify the type of structures formed. For this we computed the 3D radial distribution function $g(r)$ as shown in Fig. \ref{Figure 4}. The first peak in the $g(r)$ curves, corresponding to the average nearest-neighbor distance, is at $r = 0.8 ~\upmu \textrm{m}$. This is significantly smaller than the microgel diameter in dilute conditions ($D_{\mathrm{h}}(20 ~\degree \mathrm{C}) = 1.04 \pm 0.02 ~\upmu \textrm{m}$) as a result of the deswelling/deformation of microgels due to the dense packing. The $g(r)$ for the slow ramp contains distinct peaks indicating a highly-ordered structure. The peak positions are in agreement with the theoretical peak positions for Face-Centered Cubic (FCC) and Hexagonal Close-Packed (HCP) crystals, represented by the solid and dashed vertical lines in Fig. \ref{Figure 4}.  A mixture of FCC and HCP, or a Random Hexagonal Close Packed (RHCP) structure, is indeed the expected structure for a crystal consisting of short-range repulsive colloids. The peaks of HCP are relatively small and are only weakly present in the measured $g(r)$. In the $g(r)$ of the fast ramp the crystalline features are significantly less distinct compared to the slow ramp, indicating a less ordered final state. The $g(r)$ for the quench ramp corresponds to one expected for a glassy state, with a `split' second peak \cite{vanBlaaderen1995Real-SpaceGlasses}. Taken together, these results are in line with the common observation that cooling rate matters when a system is brought to a highly supersaturated state; when cooled slowly the particles crystallize and when cooled fast (quenched) the particles become arrested and are unable to crystallize.

\begin{figure*}[h!]
 \centering
 \includegraphics[]{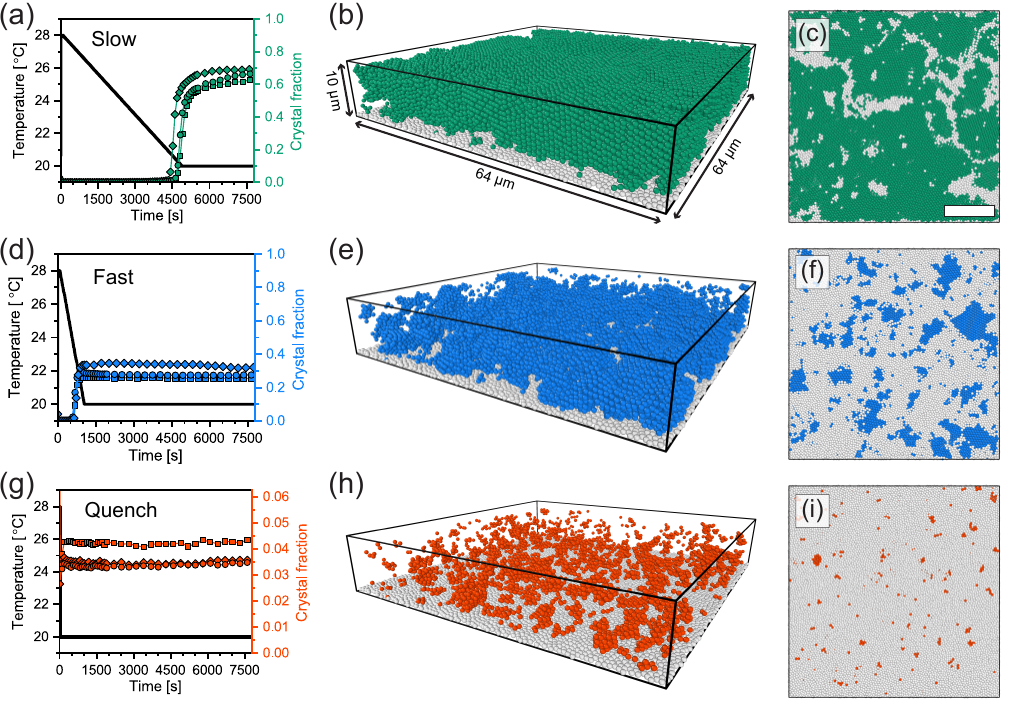}
 \caption{(a)-(d)-(g) The fraction of microgels with a crystalline structural environment over time for slow cooling (0.1 \degree C/min), fast cooling (0.5 \degree C/min) and a rapid temperature quench (>10 \degree C/min), respectively. Circles, squares and diamonds each correspond the crystal fractions determined in one run. Black lines denote the temperature profiles. (b)-(e)-(h) 3D renderings of the final state (i.e. 2 h after start of the temperature ramp) of one of the three runs. Crystalline microgels are shown as colored particles, immobile microgels on the coverslip are shown in white, and microgels not classified as crystalline are not shown. (c)-(f)-(i) Horizontal slices through 3D renderings (4-6 $\upmu \textrm{m}$ from the coverslip). Crystalline and glassy microgels are shown as colored and white particles, respectively. Scale bar is 15 $\upmu \textrm{m}$. }
 \label{Figure 5}
\end{figure*}

After having determined the effect of the cooling rate on the \textit{overall} order of the microgel system, we examined the \textit{local} ordering by determining the crystallinity on a single-particle level. From the CLSM xyzt-scans, we determined for each located microgel whether it has an ordered (i.e. crystalline) or disordered structural environment over the full time span of the experiment. To this end, we calculate the solid-liquid order parameter \cite{tenWolde1996NumericalUndercooling, Ramasubramani2020Freud:Data}
\begin{equation} \label{equation1}
    q_{l}(i, j) = \frac{\sum\limits_{m=-l}^{l} (q_{lm}(i)q_{lm}^{*}(j))}{\sqrt{\sum\limits_{m=-l}^{l}|q_{lm}(i)|^{2}} \sqrt{\sum\limits_{m=-l}^{l}|q_{lm}(j)|^{2}}}
\end{equation}
for each pair of directly neighboring microgels $i$ and $j$, where 
\begin{equation} \label{equation2}
    q_{lm}(i) = \frac{1}{N_{i}} \sum\limits_{j=1}^{N_{i}} Y_{lm}(\theta_{ij},\phi_{ij})
\end{equation}
is the bond-orientational order parameter for microgel $i$, $N_{i}$ is its number of nearest neighbors, $Y_{lm}(\theta_{ij},\phi_{ij})$ are the spherical harmonics, and $\theta_{ij}$ and $\phi_{ij}$ denote the polar and azimuthal angles describing the bond between $i$ and $j$. This method of determining the degree of order on a single-particle level has been extensively used in both simulations \cite{tenWolde1996NumericalUndercooling, Filion2010CrystalTechniques} and experiments \cite{Gasser2001Real-spaceCrystallization, Sandomirski2011HeterogeneousWall}. We determine the nearest neighbors using a Voronoi construction, and use $l=6$, which is the suitable for FCC and HCP structures \cite{Auer2005NumericalColloids} formed in microgel systems \cite{Brijitta2009RandomCrystals, Karthickeyan2017FCC-HCPCrystals, Muluneh2012DirectMicrogels}. We consider a bond between two neighboring microgels to be crystal-like if $q_{6}>0.7$ and a microgel is considered to be in a crystalline structural environment if it has six or more crystal-like bonds. 

Figures \ref{Figure 5}(a)-(d)-(g) show the temperature profiles (black lines) of the three cooling ramps along with the fraction of microgels detected as crystalline over time (colored symbols) . First of all, all three different runs for each cooling rate show similar final crystal fractions indicating the reproducibility of the measurements. For slow and fast cooling rates, we observe a sudden increase in the crystal fraction during the ramp. For the case of quenching the suspension, however, a sharp jump in crystallinity followed by a plateau is observed, indicating the microgels become completely arrested within a single time step of the measurement ($\sim 80$ s). We find that the highest crystal fraction is obtained for the slow cooling (around 0.60-0.70). The fast cooling ramp resulted in a significantly lower crystal fraction (0.25-0.35), while quenching from 28.0 \degree C to 20.0 \degree C resulted in a nearly completely glassy phase with only a small fraction of microgels in a crystalline structural environment (0.03-0.05). Clearly, the difference in total crystallinity is, as expected, dependent on the cooling rate, with slower rates leading to higher crystallinity and quenching leading to a glass-like structure. 

Interestingly, we observe that the onset of crystallization occurs around $20.4 \pm 0.4$ \degree C when a slow cooling ramp was used, while crystallization starts around $22.9 \pm 0.3$ \degree C in case of a fast ramp. As the volume fraction of PNIPAM microgels is controlled by the temperature, this implies that crystallization starts at a higher volume fraction for the slow cooling. This observation is in contrast with the expectation that the onset of crystallization is determined by a specific volume fraction. Clearly, this is not an equilibrium process, and thus the microgel suspension enters a supercooled state during cooling. Whether the volume fraction at which crystallization starts is truly higher for the slowest cooling rate, and if so, why this is the case remains unclear at present and warrants further investigation.

To understand where crystal domains have formed and what their size is, we visualized the spatial distribution of crystallized microgels after the temperature ramps. Typical 3D renderings of the final solid state are displayed in Figs. \ref{Figure 5}(b)-(e)-(h). Here, only the microgels classified as crystalline (colored) and those stuck to coverslip (white) are shown. In addition, horizontal slices through the renderings are given Figs. \ref{Figure 5}(c)-(f)-(i), in which we do show the disordered (glassy) microgels in white. Clearly, large ordered domains have formed in case of slow cooling (Figs. \ref{Figure 5}(b) and (c)): crystalline domains are separated by grain boundaries, seen as small `channels' of disordered microgels. For the case of fast cooling (Figs. \ref{Figure 5}(e) and (f)), we see small crystalline domains surrounded by disordered particles. The quenched suspension (Figs. \ref{Figure 5}(h) and (i)) only shows very small crystalline domains. We should note that even for a completely disordered structure, due to the `random' placement of particles, we expect some individual or small clusters of microgels to be classified as crystalline. From these 3D renderings we conclude that for the slowest cooling rate, a polycrystalline structure forms, indicating that nucleation of crystalline domains starts at many points in the field-of-view. In addition, for the faster cooling rate, the nucleation of crystalline domains can occur but their growth is inhibited by the rapid increase in volume fraction, while the quench prevents any formation of crystal domains of substantial size. 

\begin{figure}[t!]
 \centering
 \includegraphics[]{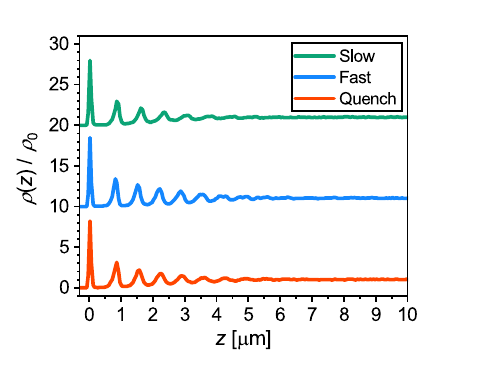}
 \caption{Normalised microgel density in the final solid phase as a function of distance $z$ from the coverslip for each cooling ramp.}
 \label{Figure 6}
\end{figure}

\begin{figure*}[h]
 \centering
 \includegraphics[]{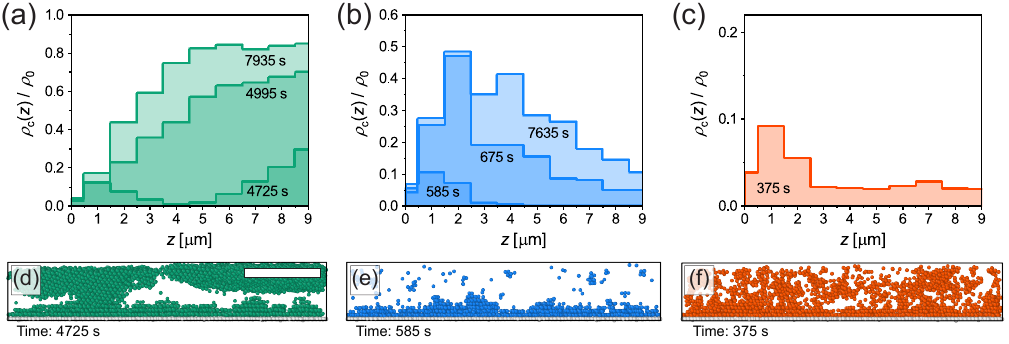}
 \caption{(a)-(b)-(c) Density of crystalline microgels as function of distance $z$ from coverslip at different points in time for slow, fast and quench cooling. Data is obtained from single runs corresponding to the circles in Figs. \ref{Figure 6}(a)-(b)-(c). Note that the scales on the vertical axes are different. (d)-(e)-(f) Snapshots (3D renderings, side view) of the crystalline microgels corresponding to the first time point given in (a)-(b)-(c). Immobile microgels on the coverslip are shown in white. Scale bar is 15 $\upmu \textrm{m}$.}
 \label{Figure 7}
\end{figure*}

\subsection{Effect of wall on crystal formation} \label{Effect of the wall}
To explore the direct effect of a flat wall --- the coverslip in our experiments --- on the crystal formation, we first investigate its influence on the microgel distribution after the three cooling ramps. To this end, we plot the microgel density relative to the mean density $\rho(z)/\rho_0$ as a function of the distance $z$ from the coverslip in Fig. \ref{Figure 6}. For all three cooling rates the density profiles show distinct peaks, where the peak at $z = 0 ~\upmu \textrm{m}$ corresponds to the immobile microgels at the coverslip. The separation between the peaks is approximately $0.7 ~\upmu \textrm{m}$, slightly smaller than the average inter-particle spacing  (i.e. positions of first peaks in Fig. \ref{Figure 4}). Evidently, the peaks in the density profiles show that a `layering' of microgels is present near the coverslip, and the decay in peak intensity with $z$ in all curves indicates that the layering becomes less pronounced the further from the wall. Particle layering next to the wall is not unexpected as even in a liquid such layering can be observed \cite{VanWinkle1988LayeringWall}. However, that a mostly similar layering is seen for all three cooling rates is surprising as large differences in the structures were observed in the $g(r)$ (Fig. \ref{Figure 4}). Upon closer inspection, the peaks in the case of slow cooling appear to decay over a slightly shorter distance than for the other two cooling rates. This can be explained by the fact that the highest crystal fraction is obtained after the slow cooling ramp. Crystal domains that are formed in the bulk, via homogeneous nucleation, will have a random orientation. As a result, the crystal layers in these domains are generally not parallel to the coverslip, explaining the slightly less distinct peaks in the density profile after slow cooling. Since the crystal fraction is much lower after the fast and quench cooling, the effect of the orientation of crystal planes does not play a major role in those density profiles.

To shed more light on the crystal nucleation mechanism near the wall, we studied the normalised density of crystalline microgels $\rho_{\mathrm{c}}(z)/\rho_0$ as function of distance $z$ from the coverslip during the nucleation process. Figures \ref{Figure 7}(a)-(b)-(c) show the extracted density profiles of the crystalline microgels for the three cooling rates; slow, fast and quench cooling, respectively. Here, a bin width of $1 ~\mu m$ is used for clarity. For slow cooling in Fig. \ref{Figure 7}(a), at $t = 4725$ s, we observe the formation of the crystal domains both on the coverslip ($z = 0 - 4 ~\upmu \textrm{m}$) and in the "bulk" ($z > 6 ~\upmu \textrm{m}$) of the imaged volume, and little crystalline microgels in the intermediate region. Figure \ref{Figure 7}(d) depicts a side view rendering in the full imaged volume, clearly showing the two different regions containing crystalline microgels. As time progresses, we see that these crystal domains come together, resulting in a structure that has a higher final crystal fraction in the bulk than near the coverslip (Fig. \ref{Figure 7}(a), at $t = 7935$ s). It appears that, for the slow cooling rate, crystals form both via heterogeneous nucleation on the wall and homogeneous nucleation in the bulk. 

For fast cooling in Fig. \ref{Figure 7}(b), we observe the formation of crystal domains only above the wall at $t = 585$ s until the end at $t = 7635$ s and no large crystal grains growing in the bulk. The presence of only crystal domains on the wall is also clearly visible in the side view rendering in Fig. \ref{Figure 7}(e). Therefore, for the fast cooling rate we conclude that crystallization is dominated by heterogeneous nucleation. 

In case of the quench in Fig. \ref{Figure 7}(c), crystallization occurred within a single frame and did not change significantly afterwards, since the microgels formed an arrested state. Therefore, only the the density profile of crystalline microgels is given at $t = 375$ s (shortly after the quench). This quench profile shows that the fraction of crystalline microgels is slightly higher near the coverslip than in the bulk, see also Fig. \ref{Figure 7}(f), again indicating the presence of heterogeneous nucleation.  

We recall that the structure of adsorbed microgels at the wall is disordered, and therefore to some extend replicates the structure of the fluid, which influences the degree of heterogeneous nucleation in our system compared to a perfectly flat wall \cite{Espinosa2019HeterogeneousSpheres}. Additionally, it should be noted here that the disordered layer of adsorbed microgels affects the identification of crystalline microgels. This results in a relatively low fraction of microgels identified as crystal near the coverslip, around $z = 0 - 1 ~\upmu \textrm{m}$, which is most apparent in Fig. \ref{Figure 7}(b). Overall, however, our results certainly illustrate that the cooling rate determines which nucleation mechanism is dominant and that it influences the final structure of the microgel suspension near the wall.

%\subsection{Crystal allignment to coverslip}
%
\begin{figure*}[h!]
 \centering
 \includegraphics[]{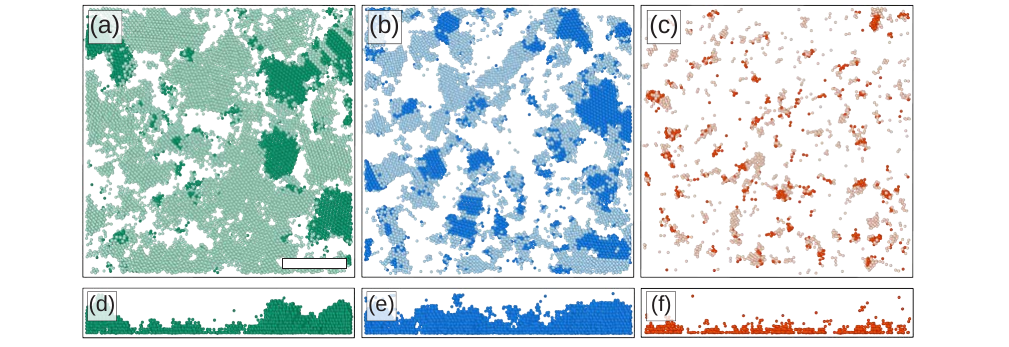}
 \caption{(a)-(b)-(c) Slices of 3D renderings (0-4 \textmu m) after the slow, fast and quench cooling ramps, where only microgels identified as FCC and HCP are shown. Dark colored microgels are in a structural environment that have an hexagonally-packed plane aligned with the coverslip ($A>0.99$). Data corresponds to the squares in Figs. \ref{Figure 6}(a)-(b)-(c), 2 h after start of the ramps.  Scale bar is 15 $\upmu \textrm{m}$. (d)-(e)-(f) Side view of microgels with $A>0.99$; other microgels are omitted.}
 \label{Figure 8}
\end{figure*}
The density profiles displayed in Fig. \ref{Figure 6} revealed the layering of microgels near the coverslip, which is already a strong indication that the wall also has an effect on the orientation of the crystal grains. It is well known that, in case of heterogeneous nucleation, colloidal particles forming an RHCP structure typically align the hexagonally-packed planes with the wall \cite{Auer2003LineColloids, Sandomirski2014HeterogeneousWalls}. To investigate this alignment after the three cooling ramps, we determined the local crystal orientation on a single-particle level relative to the wall. As a first step, we used \textit{Polyhedral Template Matching} (PTM) \cite{Stukowski2010VisualizationTool, Larsen2016RobustMatching} to determine whether the local structure around a microgel corresponds to an FCC or HCP structure. PTM identifies local structures by matching the positions of microgels and their nearest neighbors to crystal structure templates. The algorithm requires a cutoff value for the level of similarity (Root-Mean-Square Deviation, or RMSD) between the local structure and the template, for which we use RMSD = 0.2 (we use all microgels for identifying FCC and HCP structures with PTM, but only use those that are also classified as crystalline using the method described in Section \ref{Effect of cooling rate} in the subsequent analysis). With the PTM analysis we find indeed that the crystalline domains contain a mixture of FCC and HCP structures. A horizontal slice showing the typical distribution of FCC and HCP in a crystal obtained after the slow cooling ramp is given in Supplementary Fig. 5 in the ESI\dag. 

Importantly, PTM also allows us to determine the local orientation of each nearest-neighbor cluster (i.e. a microgel and its direct nearest neighbors) assigned to an FCC or HCP structure. To determine to which extend a crystal plane is aligned with the wall, we calculate the correlation 
\begin{equation} \label{equation3}
    A = (\textbf{n}_{\mathrm{c}} \cdot \textbf{n}_{\mathrm{w}})^{2},
\end{equation}
where $\textbf{n}_{\mathrm{c}}$ and $\textbf{n}_{\mathrm{w}}$ are the (unit) normal vectors of a crystal plane and the wall, respectively. Hence, $A = 1$ indicates that the crystal plane is parallel to the wall, while $A$ becomes lower for larger angles between the crystal plane and the wall. For FCC, the hexagonally-packed planes correspond to the $\{1,1,1\}$ family of planes, and in HCP it is the $(0,0,0,1)$ plane. To measure the alignment of these planes with the coverslip, we calculate the corresponding $A$ for each FCC/HCP microgel (since FCC has four perpendicular $\{1,1,1\}$ planes, we take the plane with the highest $A$, i.e. the most aligned with the coverslip). In Fig. \ref{Figure 8}(a)-(b)-(c), we show renderings of all microgels in a crystalline domain close to the wall after completion of the slow, fast and quench cooling ramps, respectively. Here microgels with $A>0.99$ (or, in other words, an alignment within $5.7 \degree$) are shown as dark colored particles. In addition, side views of the renderings are given in Fig. \ref{Figure 8}(d)-(e)-(f). For the three cooling rates, we observe that several crystalline domains possess hexagonally-packed planes aligned with the coverslip, with the quench showing only very small domains as the crystal fraction is very low. Interestingly, there appears to be a difference in the alignment of hexagonally-packed planes between the fast cooling and slow cooling rates. For the fast cooling, a relatively large fraction of crystalline domains have an alignment of a hexagonally-packed plane with the coverslip (Fig. \ref{Figure 8}(b)), while for the slow cooling most crystalline microgel domains have a different alignment (Fig. \ref{Figure 8}(a)).

\begin{figure}[t!]
 \centering
 \includegraphics[]{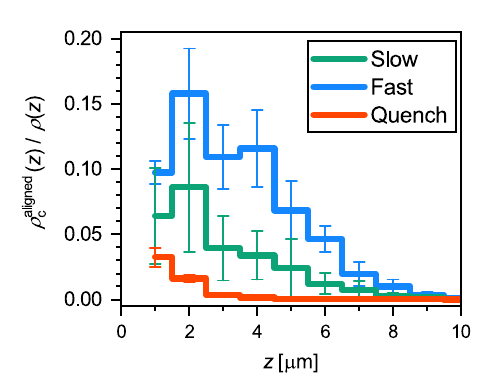}
 \caption{Fraction of microgels that are in a crystalline structural environment (FCC or HCP) and have an hexagonally-packed plane aligned with the coverslip ($A>0.99$), as function of distance $z$ from the coverslip. Results are an average of three runs for each ramp rate.}
 \label{Figure 9}
\end{figure}

To quantitatively compare the overall alignment of the hexagonally-packed layers after the three cooling ramps, we plot the fraction of crystalline microgels with $A>0.99$ as function of distance $z$ from the coverslip in Fig. \ref{Figure 9}. As expected, we find that the fraction of aligned crystals is highest near the coverslip for all three cooling rates. Confirming our previous observation of the difference in alignment of crystal domains, we find that the fraction of microgels in an aligned hexagonally-packed layer is higher after the fast cooling ramps than after the slow ramps, even though the fraction of crystalline microgels is much greater after the slow ramps (Figs. \ref{Figure 5}(a)-(d)). The lower alignment of the crystalline domains close to the wall obtained with a slow cooling rate appears to be due to a competition of the two nucleation mechanisms we have observed before (Figs. \ref{Figure 7}(a)-(d) versus Figs. \ref{Figure 7}(b)-(e)). For the slow cooling rate, the crystal domains that have a hexagonally-packed plane aligned with the coverslip originate from heterogeneous nucleation. It seems that their growth is hindered by the growing crystal domains that originate from homogeneous nucleation in the bulk that possess a different orientation. This competition between the growth of aligned and misaligned crystal domains does not play a significant role in the fast cooling ramps, as most crystal domains originate from heterogeneous nucleation events at the coverslip. Clearly, the cooling rate is an important parameter that controls crystal orientation near a wall.

% }

% Conclusions
% {
\section{Conclusions}
In summary, we have investigated the influence of cooling rate on the crystallization of a dense suspension of soft, thermoresponsive microgels near a wall. By employing microgels with a core-shell morphology, and fluorescently labeling the core only, we were able to image the microgel positions even at a high effective volume fraction on a single-particle level in 3D using CLSM. We visualized the structural evolution during the phase transition from fluid to solid induced by a temperature change from 28 to 20 \degree C with different cooling rates: slow (0.1 \degree C/min), fast (0.5 \degree C/min) and quench (>10 \degree C/min), near the coverslip covered by an adsorbed layer of disordered microgels. 

We find, as expected, that the cooling rate influences the degree of crystallization, with slow cooling leading to the highest crystal fraction and the rapid temperature quench leading to the formation of a glass-like structure. By visualizing the crystal growth in 3D, we revealed that the cooling rate combined with the presence of the wall controls the type of crystal nucleation that occurs, and this, in turn, influences the orientation of crystal domains in the region near the wall. Both homogeneous nucleation and heterogeneous nucleation occur when the microgel suspension is cooled slowly, while heterogeneous nucleation is dominant for fast cooling. For slow cooling, the growth of crystal domains with an arbitrary orientation that originate from nucleation in the bulk appear to limit the growth of domains originating from heterogeneous nucleation that have a hexagonally-packed plane aligned with the wall.

The findings presented here highlight the importance of control over cooling rate and the effects of the wall in microgel studies on phase transitions. With the structural analysis presented in this paper, we have revealed the interesting interplay between cooling rate and wall effects on the final structure of dense microgel suspensions. These results contribute to a comprehensive understanding of the role of structural details on fundamental physical phenomena in general, such as crystallization, melting, and the glass transition. In addition, these insights can assist in the development of potential applications of microgel-based materials based on its unique stimuli-responsive properties.

% Other
% {
\section*{Author Contributions} % https://credit.niso.org/
Conceptualization - MPMS, JMM; formal analysis - MPMS, TWJV; investigation - MPMS, TWJV, TCMS; methodology - MPMS, TWJV, TCMS, JMM; supervision - JMM; visualization - MPMS, TWJV; writing (original draft) - MPMS; writing (review \& editing) - MPMS, JMM. All authors read and agreed on the final text of the paper.

\section*{Conflicts of interest}
There are no conflicts to declare.

\section*{Acknowledgements}
We thank C. Storm for fruitful discussions. JMM acknowledges financial support from the Netherlands Organization for Scientific Research (NWO) (016.Veni.192.119).
% }

%%%END OF MAIN TEXT%%%


\balance


%%%REFERENCES%%%
%\bibliography{references} %You need to replace "rsc" on this line with the name of your .bib file
\input{output_MAIN.bbl}

\bibliographystyle{rsc} %the RSC's .bst file

\end{document}

%%% Style
% {
\pagestyle{fancy}
\thispagestyle{plain}
\fancypagestyle{plain}{
%%%HEADER%%%
\renewcommand{\headrulewidth}{0pt}
}
%%%END OF HEADER%%%

%%%PAGE SETUP - Please do not change any commands within this section%%%
\makeFNbottom
\makeatletter
\renewcommand\LARGE{\@setfontsize\LARGE{15pt}{17}}
\renewcommand\Large{\@setfontsize\Large{12pt}{14}}
\renewcommand\large{\@setfontsize\large{10pt}{12}}
\renewcommand\footnotesize{\@setfontsize\footnotesize{7pt}{10}}
\makeatother

\renewcommand{\thefootnote}{\fnsymbol{footnote}}
\renewcommand\footnoterule{\vspace*{1pt}% 
\color{cream}\hrule width 3.5in height 0.4pt \color{black}\vspace*{5pt}} 
\setcounter{secnumdepth}{5}

\makeatletter 
\renewcommand\@biblabel[1]{#1}            
\renewcommand\@makefntext[1]% 
{\noindent\makebox[0pt][r]{\@thefnmark\,}#1}
\makeatother 
\renewcommand{\figurename}{\small{Supplementary Fig.}~}
\sectionfont{\sffamily\Large}
\subsectionfont{\normalsize}
\subsubsectionfont{\bf}
\setstretch{1.125} %In particular, please do not alter this line.
\setlength{\skip\footins}{0.8cm}
\setlength{\footnotesep}{0.25cm}
\setlength{\jot}{10pt}
\titlespacing*{\section}{0pt}{4pt}{4pt}
\titlespacing*{\subsection}{0pt}{15pt}{1pt}
%%%END OF PAGE SETUP%%%
\renewcommand{\thesection}{S\arabic{section}}

%%%FOOTER%%%
\fancyfoot{}
%\fancyfoot[LO,RE]{\vspace{-7.1pt}\includegraphics[height=9pt]{head_foot/LF}}
%\fancyfoot[CO]{\vspace{-7.1pt}\hspace{13.2cm}\includegraphics{head_foot/RF}}
%\fancyfoot[CE]{\vspace{-7.2pt}\hspace{-14.2cm}\includegraphics{head_foot/RF}}
\fancyfoot[RO]{\footnotesize{\sffamily{1--\pageref{LastPage} ~\textbar  \hspace{2pt}\thepage}}}
\fancyfoot[LE]{\footnotesize{\sffamily{\thepage~\textbar\hspace{3.45cm} 1--\pageref{LastPage}}}}
\fancyhead{}
\renewcommand{\headrulewidth}{0pt} 
\renewcommand{\footrulewidth}{0pt}
\setlength{\arrayrulewidth}{1pt}
\setlength{\columnsep}{6.5mm}
\setlength\bibsep{1pt}
%%%END OF FOOTER%%%

%%%FIGURE SETUP - please do not change any commands within this section%%%
\makeatletter 
\newlength{\figrulesep} 
\setlength{\figrulesep}{0.5\textfloatsep} 

\newcommand{\topfigrule}{\vspace*{-1pt}% 
\noindent{\color{cream}\rule[-\figrulesep]{\columnwidth}{1.5pt}} }

\newcommand{\botfigrule}{\vspace*{-2pt}% 
\noindent{\color{cream}\rule[\figrulesep]{\columnwidth}{1.5pt}} }

\newcommand{\dblfigrule}{\vspace*{-1pt}% 
\noindent{\color{cream}\rule[-\figrulesep]{\textwidth}{1.5pt}} }

\makeatother
%%%END OF FIGURE SETUP%%%
% }

%%%EQUATION SETUP%%%
\makeatletter
\renewcommand\theequation{S\@arabic\c@equation}
\makeatother
%%%END OF EQUATION SETUP%%%

\begin{titlepage}
   \begin{center}
       \vspace*{1cm}

       \textbf{\Huge 3D Visualization Reveals Cooling Rate Dependent Crystallization near a Wall in Dense Microgel Systems}

       \vspace{0.5cm}
       \textit{\Large ELECTRONIC SUPPLEMENTARY INFORMATION}
            
       \vspace{1.5cm}

       \textbf{\Large M.P.M. Schelling, T.W.J. Verouden, T.C.M. Stevens, J.-M. Meijer}

       \vspace{1.5cm}
            
        {\Large Department of Applied Physics and Science Education, Eindhoven University of Technology, \\P.O. Box 513, 5600 MB Eindhoven, The Netherlands}\\
        \vspace{0.4cm}
        {\Large Institute for Complex Molecular Systems, Eindhoven University of Technology, \\P.O. Box 513, 5600 MB Eindhoven, The Netherlands}
            
   \end{center}
\end{titlepage}

\section{Microgel synthesis and characterization}
We used the synthesis procedure reported in [J. Appel et al., \textit{Part. Part. Syst. Charact.}, \textbf{32}, 764–770 (2015)]\cite{Appel2015TemperatureSuspensions} to prepare PNIPAM microgels containing a fluorescent core. All chemicals were used as received from the suppliers. For the core synthesis 8.93 g 2,2,2-trifluoroethyl methacrylate (TCI Chemicals, 98\%), 0.994 g \textit{N}-isopropyl acrylamide (NIPAM, Merck, 97\%), 16.9 mg sodium dodecyl sulfate (Merck, 98.5\%), $\sim 10$ mg Pyrromethene 546 (Merck) and 33 mL water (purified using a MilliQ\textsuperscript{\textregistered} Direct 8 system, 18.2 M$\Omega\cdot$cm) were placed in a 100 mL round-bottom flask. The mixture was bubbled with $\mathrm{N_{2}}$ for 30 min while being heated to 75 \degree C and stirred at 400 RPM (Fisherbrand Oval PTFE Stir Bar, 25x12mm). The polymerization was initiated by injection of 2.5 mL a 20 mg/mL potassium persulfate (KPS, Merck, 99\%) solution, and left for 4 hours (75 \degree C, 400 RPM stirring). Finally, the latex suspension was filtered to remove any coagulum. 

Microgel shells were obtained via seeded precipitation polymerization. 1.195 g NIPAM, 17.4 mg \textit{N},\textit{N'}-methylenebisacrylamide (Merck, 99\%), 66.7 $~\upmu \textrm{L}$ methacrylic acid (Merck, 99\%), 1 mL core suspension (21.5 wt\%) and 50 mL water were placed into a 100 mL round-bottom flask. Again, the mixture was bubbled with $\mathrm{N_{2}}$ for 30 min while being heated to 75 \degree C and stirred at 400 RPM. 1 mL KPS solution (51 mg/mL water) was added to start the polymerization, and the mixture was left for 4 hours (75 \degree C, 400 RPM stirring). The microgel suspension was filtered, and cleaned by repeated centrifugation and redispersion in MilliQ water.

The hydrodynamic size of the microgels was determined with dynamic light scattering using an Anton Paar Litesizer 500 (658 nm, 90 \degree). Further, the $\zeta$-potential was determined with electrophoretic light scattering using the same device. Samples were prepared at 0.02 wt\% in 10 mM NaCl solution. The results of the measurements are given in Supplementary Fig. \ref{Figure S1}.

\begin{figure}[h]
\centering
  \includegraphics[]{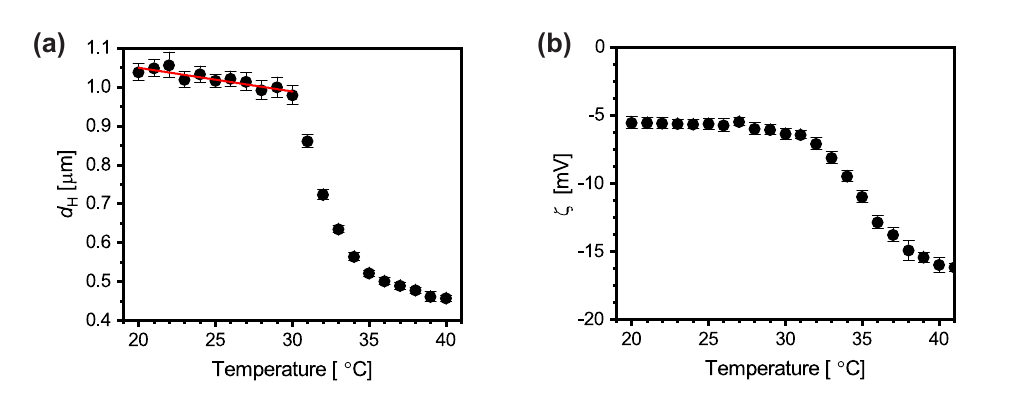}
  \caption{(a) Hydrodynamic diameter $d_\mathrm{H}$ of the core-shell microgels as function of temperature. Error bars denote the 95 \% confidence interval computed from the standard error of the mean (typically from 8 measurements). The red line is a linear fit to indicate the small size change before the Volume Phase Transition. (b) $\zeta$-potential the core-shell microgels as function of temperature. Error bars denote one standard deviation.}
  \label{Figure S1}
\end{figure}

\newpage
\section{Temperature-controlled CLSM sample cell}
A schematic illustration of the sample cell used for the experiments is given in Supplementary Fig. \ref{Figure S2}. The glass ring has an inner diameter of 5 mm and height of approximately 2 mm. The VAHEAT substrate, glass ring and coverslips are glued together using UV glue (Norland Optical Adhesive 68). The objective collar is kept at a constant temperature of 15 \degree C to allow for fast cooling close to room temperature.

\begin{figure}[h]
\centering
  \includegraphics[width = 0.5\textwidth]{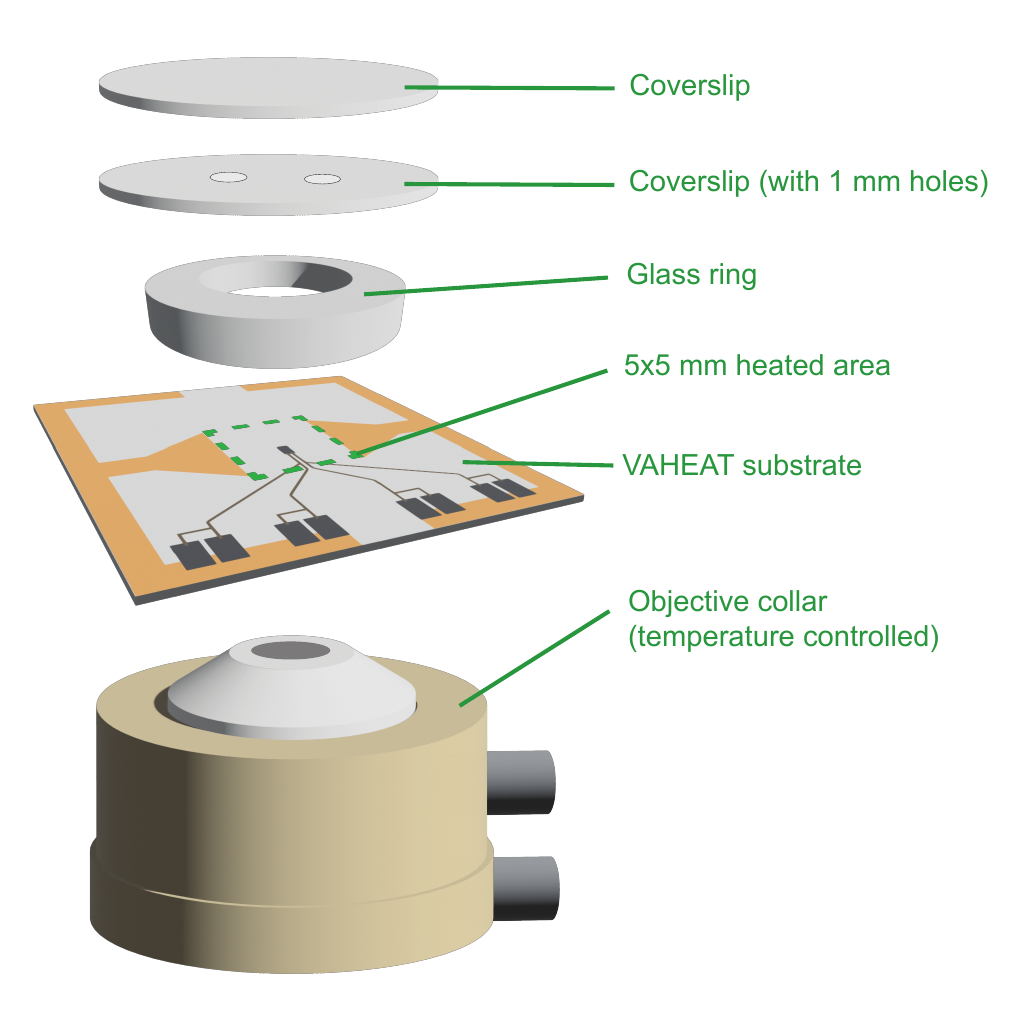}
  \caption{Schematic illustration of the sample cell used for the temperature-controlled CLSM experiments.}
  \label{Figure S2}
\end{figure}

\newpage
\section{Melting point of the dense microgel suspension}
\begin{figure}[h]
\centering
  \includegraphics{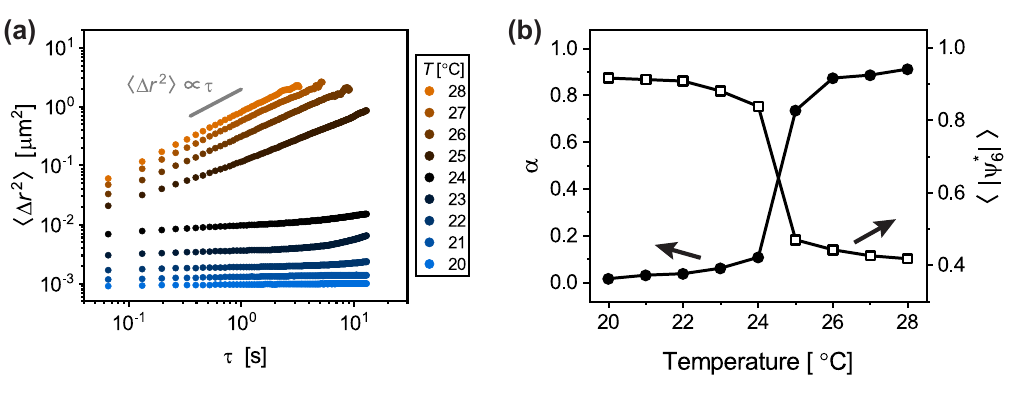}
  \caption{(a) Mean squared displacement $\langle \Delta r^{2} \rangle$ as function of lag time $\tau$ between 20.0 and 28.0 \degree C. (b) Left axis, solid circles: $\alpha$ obtained by fitting the data in (a) using Eq. (\ref{equation_s2}) for $\tau < 3$ s.  Right axis, open squares: $\langle |\psi_{6}^{*}| \rangle$ as function of temperature.}
  \label{Figure S3}
\end{figure}

To find the melting temperature of the microgel suspension, the sample was heated manually from 20.0 to 28.0 \degree C in steps of 1.0 \degree C. Before the heating experiment, the sample was left at room temperature for 2 months, resulting in a crystallized microgel suspension with hexagonally-packed planes aligned with the coverslip. At each temperature, CLSM xyt-scans (120 s, 15.3 fps) were acquired at a distance of approximately $3 ~\upmu \textrm{m}$ from the coverslip (corresponding to the fourth layer including the microgels stuck to the coverslip). The xyt-scans were acquired in a 32x32 $~\upmu \textrm{m}$ ($0.063~\upmu \textrm{m/px}$) field-of-view containing around $2 \cdot 10^{3}$ microgels. From the located microgels, the mean squared displacement 

\begin{equation} \label{equation_s1}
    \langle \Delta r^{2} \rangle (\tau) = \langle | \mathbf{r}(t + \tau) - \mathbf{r}(t) |^{2} \rangle
\end{equation}
was calculated, where $\mathbf{r}=(x,y)$ is the microgel position, $\tau$ is the lag time, and $\langle ... \rangle$ denotes the average over time $t$ and all detected microgels. Supplementary Fig. \ref{Figure S3}(a) shows $\langle \Delta r^{2} \rangle (\tau)$ for temperatures between 20.0 to 28.0 \degree C. At low temperatures, $\langle \Delta r^{2} \rangle$ displays a plateau, as the microgels remain on their lattice site. At elevated temperatures, $\langle \Delta r^{2} \rangle$ increases with $\tau$ due to long-time diffusion of the microgels, indicating a fluid phase. This transition from crystal to fluid can be clearly visualized by fitting a power law

\begin{equation} \label{equation_s2}
    \langle \Delta r^{2} \rangle (\tau) = C \tau^{\alpha},
\end{equation}
where $C$ and $\alpha$ are fitting parameters. For Brownian diffusion $\alpha = 1$, while for subdiffusive motion  $0 < \alpha < 1$ (e.g. due to caging). The fitted values for $\alpha$ are given in Supplementary Fig. \ref{Figure S3}(b) and clearly display a transition around 24-25 \degree C from a solid with near-zero values for $\alpha$, to a fluid with values close to $\alpha = 1$. 

Furthermore, instead of considering the dynamical behavior of the microgels (i.e. their mean squared displacement), the melting point was determined by investigating the degree of order during heating from 20.0 to 28.0 \degree C. To this end, we define the weighted bond-orientational order parameter \cite{Ramasubramani2020Freud:Data, Mickel2013ShortcomingsMatter, Lotito2020PatternTechniques} for each microgel $n$ as

\begin{equation} \label{equation_s3}
    \psi_{6}^{*} (n) = \frac{1}{\sum_{m=1}^{N} w_{nm}} \cdot \sum_{m=1}^{N} w_{nm} \mathrm{e}^{6i\theta_{nm}},
\end{equation}
where $N$ is the number of nearest neighbors $m$ of microgel $n$, and $\theta_{nm}$ is the angle between the vector from microgel $n$ to nearest neighbor $m$ and a reference axis. Furthermore, $w_{nm}$ is a weight proportional to the length the Voronoi cell edge. By averaging $|\psi_{6}^{*}|$ over all detected microgels, we obtain the global bond-orientational order parameter $\langle |\psi_{6}^{*}| \rangle$, which indicates the degree of hexatic ordering. For a crystal with perfect hexagonal packing $\langle |\psi_{6}^{*}| \rangle = 1$, while $\langle |\psi_{6}^{*}| \rangle < 1$ implies that hexatic order is lost. Supplementary Fig. \ref{Figure S3}(b) shows $\langle |\psi_{6}^{*}| \rangle$ as function of temperature. Indeed, we find that $\langle |\psi_{6}^{*}| \rangle$ sharply decreases around 24-25 \degree C, indicating a transition from a hexagonally-packed crystal to a fluid. 

\newpage
\section{Sample temperature after quench}
\begin{figure}[h]
\centering
  \includegraphics{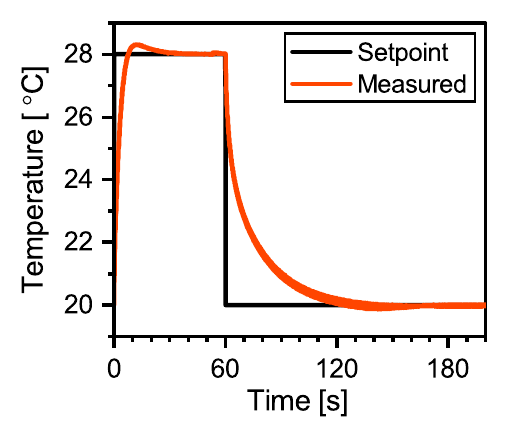}
  \caption{Comparison between the set point temperature and the measured temperatures for the case of the rapid temperature quench. The colored line consists of three overlapping lines corresponding to the three runs.}
  \label{Figure S4}
\end{figure}

\newpage
\section{Determination of HCP and FCC structures with PTM}
\begin{figure}[h]
\centering
  \includegraphics{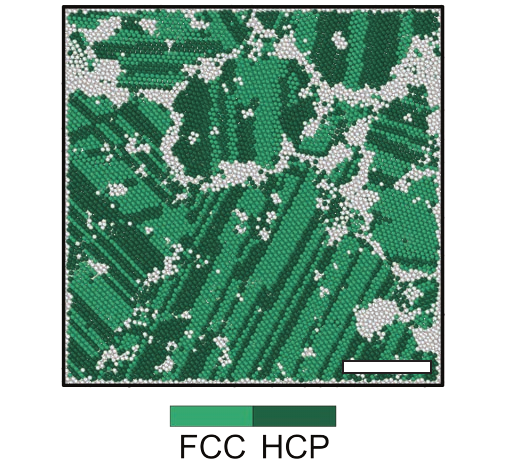}
  \caption{Slice of a 3D rendering (6-8 $\upmu \textrm{m}$ from coverslip) during the slow cooling ramp (squares in Fig. 5(a); 5535 s after the start of the ramp). Colored particles denote those identified as FCC or HCP using PTM \cite{Stukowski2010VisualizationTool, Larsen2016RobustMatching} (RMSD < 0.2); white particles are assigned to neither FCC nor HCP. Scale bar is $15 ~\upmu \textrm{m}$.}
  \label{Figure S5}
\end{figure}

\newpage
%\bibliography{references} %You need to replace "rsc" on this line with the name of your .bib file
\input{output_ESI.bbl}

\bibliographystyle{rsc}

%% file: output_MAIN.bbl
\providecommand*{\mcitethebibliography}{\thebibliography}
\csname @ifundefined\endcsname{endmcitethebibliography}
{\let\endmcitethebibliography\endthebibliography}{}

%% file: output_ESI.bbl
\providecommand*{\mcitethebibliography}{\thebibliography}
\csname @ifundefined\endcsname{endmcitethebibliography}
{\let\endmcitethebibliography\endthebibliography}{}